\documentclass[twocolumn,pra,showpacs]{revtex4}

\usepackage{dcolumn}
\usepackage{bm}

\def\etal{\emph{et~al.}}
\newcommand{\eref}[1]{(\ref{#1})}
\newcommand{\Eref}[1]{Eq.~(\ref{#1})}
\newcommand{\tref}[1]{Table~\ref{#1}}
\newcommand{\rtw}{\rightarrow}

\begin{document}

\title{Configuration interaction calculation for the isotope shift in Mg~I.}

\author{J. C. Berengut}
\email{jcb@phys.unsw.edu.au}
\affiliation{School of Physics, University of New South Wales, Sydney 2052, Australia}
\author{V. A. Dzuba}
\affiliation{School of Physics, University of New South Wales, Sydney 2052, Australia}
\author{V. V. Flambaum}
\affiliation{School of Physics, University of New South Wales, Sydney 2052, Australia}
\affiliation{Institute for Advanced Study, Einstein drive, Princeton, NJ 08540, USA}
\author{M. G. Kozlov}
\affiliation{Petersburg Nuclear Physics Institute, Gatchina, 188300, Russia}
\affiliation{Queen's University of Belfast, Belfast, BT71NN, UK}

\date{23 December 2003}

\begin{abstract}

We present an {\it ab initio} method of calculation of isotope shift
in atoms with a few valence electrons, based on the
configuration-interaction calculation of energy. The main motivation
for developing the method comes from the need to analyze whether
differences in isotope abundance in early universe can contribute to
the observed anomalies in quasar absorption spectra. The current best
explanation for these anomalies is the assumption that the fine
structure constant $\alpha$ was smaller at early epoch. We show that
we can calculate the isotope shift in magnesium with good accuracy.

\end{abstract}


\maketitle

\section{\label{sec:intro} Introduction}

The motivation for this work comes from recent studies of quasar
absorption spectra that reveal a possible change in $\alpha$ since the
early universe \cite{murphy1}. One of the possible major sources of
systematic effects in these studies is that the isotopic abundance
ratios in gas clouds in the early universe could be 
very different to those on Earth. A ``conspiracy'' of several isotopic
abundances may provide an alternative explanation for the observed 
variation in spectra \cite{murphy2}.
In order to test this possibility
it is necessary to have accurate values for the isotope shift (IS) for
the relevant atomic transitions. Experimental data is available for
only a very few of them; therefore, accurate calculations are needed
to make the most comprehensive analysis possible.

Previously we have calculated isotope shift in atoms with one valence
electron \cite{berengut}. This work represents a first step in
developing a method for the calculation of isotope shift in atoms with
more than one valence electron. 
The method used here is similar to our previous calculations of the
$\alpha$-dependence of transition frequences for ions with several
valence electrons~\cite{DFKM}. It includes Dirac-Fock calculation of the 
core and configuration interaction (CI) for the valence electrons in 
combination with the finite-field method for the perturbation.
Magnesium is one of the simplest and
well studied two-electron atoms. Because of that it is often used as a
test ground for different methods of atomic calculations. In this
paper we show that we can calculate the isotope shift of some
magnesium transitions for which experimental values are available.

\section{\label{sec:method} Method}

The isotope shifts of atomic transition frequencies come from two
sources: the finite size of the nuclear charge distribution (the
``volume'' or ``field'' shift), and the finite mass of the nucleus
(see, e.g. \cite{sobelman}). The energy shift due to recoil of the
nucleus is $(1/2M)\bm{p}_N^2~=~(1/2M)(\Sigma \bm{p}_i)^2$. 
Furthermore this ``mass shift'' is traditionally divided
into the normal mass shift (NMS) and the specific mass shift (SMS). The
normal mass shift is given by the operator $(1/2M)\Sigma \bm{p}_i^2$, 
which is easily calculated from the transition frequency. The
SMS operator is $(1/M)\Sigma_{i<j}(\bm{p}_i \cdot \bm{p}_j)$ which is
difficult to evaluate accurately.

The shift in energy of any transition in an isotope with mass number
$A'$ with respect to an isotope with mass number $A$ can be expressed
as
\begin{equation}
\label{eq:is}
\delta \nu^{A', A} = \left( k_{\rm NMS} + k_{\rm SMS} \right)  \left(
    \frac{1}{A'} - \frac{1}{A} \right) + F \delta \langle r^2 \rangle
    ^{A', A} \ ,
\end{equation}
where the normal mass shift constant is
\begin{equation}
k_{\rm NMS} = -\frac{\nu}{1823}
\end{equation}
and $\langle r^2 \rangle$ is the mean square nuclear radius. The value
1823 refers to the ratio of the atomic mass unit to the electron mass.

In this paper we develop a method for calculating the specific mass
shift $k_{\rm SMS}$ for atoms with several valence electrons.
It is worth noting that in this paper we use the
convention $\delta \nu^{A', A}~=~\nu^{A'} - \nu^{A}$.

Following our previous work on single valence electron atoms
(Ref. \cite{berengut}) we are looking for an ``all order'' method of
calculation. Again we have found that the finite-field scaling method
is very useful in this respect. The rescaled SMS operator is added to
the many-particle Hamiltonian
\begin{eqnarray}
\label{eq:H_sms}
H_{\lambda} = H_0 + \lambda H_{\rm SMS} = H_0 
+ \lambda \sum_{i<j}\bm{p}_i \cdot \bm{p}_j.
\end{eqnarray}
The eigenvalue problem for the new Hamiltonian is solved for various
$\lambda$, and then we recover the specific mass shift constant as
\begin{equation}
k_{\rm SMS} = \lim_{\lambda \rtw 0} \frac{dE}{d\lambda}.
\end{equation}
The operator \eref{eq:H_sms} has the same symmetry and structure as
the initial Hamiltonian $H_0$ (see the Appendix, Ref.~\cite{berengut}).

In this work we restrict ourselves to the frozen-core
approximation. We first solve the Dirac-Fock equations for the core
and valence electrons. Then we generate a basis set that includes the
core and valence orbitals and a number of virtual orbitals. Finally we
do the full configuration interaction calculation.

The SMS operator for the valence electrons in the frozen core
approximation can be divided into the core, one-electron and
two-electron parts:
\begin{equation}
\label{eq:H_breakdown}
H_{\rm SMS} = H_{\rm SMS}^{(0)} + H_{\rm SMS}^{(1)} + H_{\rm SMS}^{(2)}
\end{equation}
The first term in \Eref{eq:H_breakdown} corresponds to the change of
the core potential. It accounts for the change of the core orbitals
when the Dirac-Fock equations are solved for the operator
$H_{\lambda}$. The term $H_{\rm SMS}^{(1)}$ accounts for the exchange
interaction of the valence electrons with the core:
\begin{equation}
\label{eq:H_1}
\langle i | H_{\rm SMS}^{(1)} | k \rangle =
\sum_{j=1}^{N_{\text{core}}} \langle i, j | \bm{p}_1 \cdot \bm{p}_2
| j, k \rangle.
\end{equation}
The last term corresponds to the specific mass shift between the two
valence electrons, $\bm{p}_1 \cdot \bm{p}_2$.

\section{Calculation and results}

We are using the Dirac-Fock code \cite{BDT77}, which was modified 
for the Hamiltonian \eref{eq:H_sms}. The CI calculations are made with 
the help of the modification \cite{dzuba96} of the code \cite{KT87}. 
In order to study the role of the valence correlations we made three 
different calculations:

\begin{enumerate}
\item The basic one-configurational calculation for the ground
    state $^1S_0[3s^2]$ and for the $^{1,3}P_J[3s3p]$ states. All core
    orbitals and orbitals $3s$ and $3p$ are formed in the $V^{N-2}$
    approximation (i.e. by solving Dirac-Fock equations for the core).
\item Full two-electron CI for the medium size basis set
    $[8sp5d]$, which includes the orbitals $1-8s_{1/2}$, $2-8p_j$ and
    $3-5d_j$. The $3s$, $3p$ and $3d$ orbitals are solutions of the
    $V^{N-2}$ Dirac-Fock potential. The remaining virtual orbitals are
    constructed by multiplying the previous orbital of the same
    partial wave by the simple radial function and orthogonalizing
    with the other orbitals~\cite{Bv83}.  
\item Full two-electron CI for the basis
    set $[12spd9f]$. This basis set is formed by diagonalizing the
    Dirac-Fock operator on the basis set of $B$-splines and excluding
    orbitals with high energy (for a
    description of this method as applied in atomic physics, see
    e.g. Ref.~\cite{johnson}). 
\end{enumerate}
Below we refer to these calculations as small, medium, and large.
The large calculation is already very close to the saturation of the
valence CI. Here the difference between the theoretical spectrum and
experiment is mostly caused by the neglect of the core-valence
correlations. The latter were studied, for example, 
in Ref.~\cite{PKRD}. For Mg~I the typical corrections to the valence 
energies and transition frequencies were found to be of the order of 
a few percent.

\tref{tab:levelshifts} presents the resulting SMS level shift
constants, $k_{\rm SMS}$ of \Eref{eq:is}, in different
approximations. The contributions of individual terms in
\Eref{eq:H_breakdown} are given, as well as their sum. It is
interesting to note that all of the terms are large in comparison to
the total SMS. There is a large cancellation between contributions
within levels, and also between different levels. This shows that very
high accuracy is required in order to preserve the remaining SMS in
transitions.

\begin{table}

\caption{Calculations of the specific mass shift constants $k_{\rm
SMS}$ for Mg~I levels (in $ \rm GHz \cdot amu$). Individual
contributions from \Eref{eq:H_breakdown} are presented, as well as the
total.  For some levels we give medium (M) CI and one-configurational
results (S) in addition to the large (L) CI ones.}

\label{tab:levelshifts}

\begin{tabular}{lrrrrc}
\hline \hline

&\multicolumn{5}{c}{$k_{\rm SMS}$} \\ \multicolumn{1}{c}{Level}
&\multicolumn{1}{c}{(0)} &\multicolumn{1}{c}{(1)}
&\multicolumn{1}{c}{(2)} &\multicolumn{1}{c}{$\Sigma$}
&\multicolumn{1}{c}{CI} \\ 
\hline 
$^1S_0(3s^2)$   &$ 559 $&$ -883 $&$  131  $&$ -193  $& L \\ 
$\quad$"        &$ 561 $&$ -881 $&$  135  $&$ -186  $& M \\ 
$\quad$"        &$ 857 $&$-1125 $&$    0  $&$ -268  $& S\\ 
$^3S_1(3s4s)$   &$ 422 $&$ -615 $&$   44  $&$ -149  $& L \\ 
$\quad$"        &$ 431 $&$ -624 $&$   52  $&$ -142  $& M \\ 
$^1S_0(3s4s)$   &$ 415 $&$ -615 $&$   21  $&$ -179  $& L \\ 
$\quad$"        &$ 424 $&$ -630 $&$   30  $&$ -177  $& M \\ 
$^1D_2(3s3d)$   &$ 343 $&$ -616 $&$ -267  $&$ -541  $& L \\ 
$^3D_1(3s3d)$   &$ 375 $&$ -561 $&$   41  $&$ -144  $& L \\
$\quad$"        &$ 381 $&$ -571 $&$  -10  $&$ -200  $& M \\
$^3D_2(3s3d)$   &$ 375 $&$ -561 $&$   41  $&$ -144  $& L \\
$^3D_3(3s3d)$   &$ 375 $&$ -561 $&$   41  $&$ -144  $& L \\
$^3P_0^o(3s3p)$ &$ 428 $&$ -853 $&$ -144  $&$ -570  $& L \\
$^3P_1^o(3s3p)$ &$ 428 $&$ -852 $&$ -145  $&$ -569  $& L \\
$^3P_2^o(3s3p)$ &$ 428 $&$ -850 $&$ -145  $&$ -567  $& L \\ 
$\quad$"        &$ 431 $&$ -850 $&$ -142  $&$ -561  $& M \\ 
$\quad$"        &$ 759 $&$-1161 $&$ -266  $&$ -668  $& S \\ 
$^1P_1^o(3s3p)$ &$ 408 $&$ -698 $&$  329  $&$   38  $& L \\ 
$\quad$"        &$ 411 $&$ -700 $&$  341  $&$   52  $& M \\ 
$\quad$"        &$ 946 $&$-1163 $&$  265  $&$   49  $& S\\ 
$^3P_0^o(3s4p)$ &$ 402 $&$ -630 $&$   13  $&$ -215  $& L \\
$^3P_1^o(3s4p)$ &$ 402 $&$ -629 $&$   13  $&$ -215  $& L \\
$^3P_2^o(3s4p)$ &$ 402 $&$ -629 $&$   13  $&$ -214  $& L \\ 
\hline\hline
\end{tabular}
\end{table}

Comparison of the different approximations shows a strong dependence
on the size of the basis sets. We see that it is very important to 
saturate the basis as completely as possible. In some cases the SMS changes
drastically even between the medium and the large basis sets. In particular,
the difference between large and medium SMS calculation for the level 
$^3D_1(3s3d)$ is 39\%. That is mostly due to the $f$-wave contribution,
which is absent in the medium basis set. Note that the SMS operator can only mix 
orbitals with $\Delta l=1$. That is why the $f$-wave contribution is more
important for the levels of the configuration $3s3d$. On the other hand, 
for the same reason, the contribution of the higher partial waves to the 
considered levels is suppressed.

Analysis of \tref{tab:levelshifts} shows that valence correlations tend to
decrease the contributions of the first two terms of the SMS operator. The 
third (two-particle) term of the SMS operator is generally not screened. On 
the contrary, for some levels the two particle contribution grows with the 
size of the basis set. Note that the final value of the two-particle 
contribution to the ground state SMS is of the same order as the other 
contributions, as it is for most other states, while in the one-configurational approximation it is zero.

In \tref{tab:comparisons} we compare the results of our calculation
with experiment for SMS in transitions between $^{26}{\rm Mg}$ and
$^{24}{\rm Mg}$. Also presented for comparison are the results of
Veseth (Ref.~\cite{veseth}). That paper used non-relativistic
many-body perturbation theory within the algebraic approximation to
calculate the isotope shift to third order for some transitions. 


\begin{table*}[htb]

\caption{Comparison with experiment of the SMS for several transitions
(in MHz) between $^{26}{\rm Mg}$ and $^{24}{\rm Mg}$. Also presented
are the results of Ref.~\cite{veseth} for a theoretical comparison. We
have assumed that the field shift is negligible. }

\label{tab:comparisons}

\begin{tabular}{lclcrrrrr}
\hline \hline && &\multicolumn{1}{c}{$\lambda$} &\multicolumn{1}{c}{IS
(expt.)}  &\multicolumn{1}{c}{NMS} &\multicolumn{3}{c}{SMS} \\
\multicolumn{3}{c}{Transition} &\multicolumn{1}{c}{(\AA)}
&&&\multicolumn{1}{c}{Expt.}  &\multicolumn{1}{c}{Present}
&\multicolumn{1}{c}{Ref.~\cite{veseth}} \\ \hline
$^1S_0(3s^2) $&$ \rtw $&$ ^3P_1^o(3s3p)$ & 4572 & 2683(0)
    \footnotemark[1] & 1153 & 1530 & 1205   & 1378 \\ 
$^1S_0(3s^2) $&$
    \rtw $&$ ^1P_1^o(3s3p)$ & 2853 & 1412(21)\footnotemark[2] & 1848 &
    -436 & -740   & \\
$\quad$"    &        &  $\quad$"       &      &
    1390(31)\footnotemark[3] &      &  -458 &       & \\
$^3P_0^o(3s3p) $&$ \rtw $&$ ^3S_1(3s4s)$ & 5169 & -396(6)
    \footnotemark[4] & 1020 & -1416 & -1349 & \\
$^3P_1^o(3s3p) $&$
    \rtw $&$ ^3S_1(3s4s)$ & 5174 & -390(5) \footnotemark[4] & 1019 &
    -1409 & -1346 & \\
$^3P_2^o(3s3p) $&$ \rtw $&$ ^3S_1(3s4s)$ & 5185
    & -390(7) \footnotemark[4] & 1017 & -1407 & -1340 & \\
$^3P_1^o(3s3p) $&$ \rtw $&$ ^3P_0(3p^2)$ & 2782
    & 1810(80) \footnotemark[5] & 1895 & -85 & -487 & \\
$^3P_0^o(3s3p) $&$ \rtw $&$ ^3D_1(3s3d)$ & 3830 &
    60(15)\footnotemark[2] & 1376 & -1316 & -1365 & -1269 \\
$^3P_1^o(3s3p) $&$ \rtw $&$ ^3D_{1,2}(3s3d)$ & 3833 & 61(3)
    \footnotemark[2] & 1375 & -1314 & -1362 & \\
$^3P_2^o(3s3p) $&$
    \rtw $&$ ^3D_{1,2,3}(3s3d)$ & 3839 & 58(4) \footnotemark[2] & 1373
    & -1315 & -1356 & \\
$^3P_1^o(3s3p) $&$ \rtw $&$ ^3D_1(3s4d)$ & 3094
    & 420(20) \footnotemark[5] & 1704 & -1284 & -1375 & \\
$^1P_1^o(3s3p) $&$ \rtw $&$ ^1D_2(3s4d)$ & 5530
    & 2107(15) \footnotemark[4] & 953 & 1154 & 1224 & \\

\hline\hline
\end{tabular}
\footnotetext[1]{Sterr \etal, 1992 \cite{sterr}}
\footnotetext[2]{Hallstadius, 1979 \cite{hallstadius79}}
\footnotetext[3]{Le Boiteux \etal, 1988 \cite{boiteux}}
\footnotetext[4]{Hallstadius and Hansen, 1978 \cite{hallstadius78}}
\footnotetext[5]{Novero \etal, 1992 \cite{novero}}
\end{table*}


We have also applied the finite-field scaling method to calculate the field
shift in Mg. By following the definitions for field shift given in
Ref.~\cite{berengut}, and using the same approximations for the CI calculation,
we found that it was less than 2\% of the normal mass shift for all relevant
transitions. The field shift is smaller than omissions in the mass shift calculation, 
notably the core-valence correlations; thus for simplicity we have neglected the field 
shift from our analysis in this paper.

Core-valence correlations have been studied for the one-electron atoms in 
Ref.~\cite{berengut, js01} and shown to be quite noticeable. They can explain
the difference between our calculations and the experiment in 
\tref{tab:comparisons}. Core-valence correlations are usually 
more important for the ground state than for excited states. That may
be the reason why the largest discrepancy with the experiment is for
the transitions from the ground state.

\section{Conclusion}

We have presented a method for the calculation of the isotope-shift in
many-electron atoms using the CI for the valence electrons in
combination with the finite-field method, and tested the method in
magnesium. The agreement was found to be quite good for all
transitions. Even for the transitions from the ground state $^1S_0$ to the $J=1$ levels
of the configuration $3s3p$, where the error is largest, it constitutes 
about 20\% of the total IS.
In particular, for the purposes of resolving systematic
errors in the search for $\alpha$-variation (Ref.~\cite{murphy1, murphy2}), 
such accuracy is high enough.

Further work on magnesium could include core-correlations, using the
extensions to CI outlined in Ref.~\cite{dzuba96}. We have decided not
to do this, however, because we wanted a general method for calculating
IS in many electron atoms. The method of including core-valence correlations
in the valence CI with the help of the effective Hamiltonian has proven to be
very effective for atoms with two or three valence electrons, but
becomes less reliable for atoms with more than three valence
electrons. Unfortunately, most of the ions of astrophysical interest
have many electrons in the open shells. For such ions valence correlations
are the most important ones and we plan to use this technique to 
calculate isotope shift for the transitions that were used to detect 
variation of $\alpha$. That will provide stringent limits on the size of 
the systematic error due to variation in isotope abundance.

\section{Acknowledgments}

This work is supported by the Australian Research Council,
Gordon Godfrey fund, and Russian Foundation for Basic Research,
grant No.~02-0216387. V.F. is grateful to the Institute
 for Advanced Study and Monell foundation for hospitality
 and support. M.K. is grateful to Queen's University of Belfast
 for hospitality.

\end{document}